# Atomically Resolved Surface Structure of SrTiO$_3$(001) Thin Films Grown in Step-Flow Mode by Pulsed Laser Deposition

Katsuya Iwaya[1*], Takeo Ohsawa[1], Ryota Shimizu[1,2], Tomihiro Hashizume[1,3,4], Taro Hitosugi[1]

[1]WPI-Advanced Institute for Materials Research, Tohoku University, Sendai 980-8577, Japan

[2]Department of Chemistry, The University of Tokyo, Bunkyo, Tokyo 113-0033, Japan

[3]Advanced Research Laboratory, Hitachi, Ltd., Hatoyama, Saitama 350-0395, Japan

[4]Department of Physics, Tokyo Institute of Technology, Meguro, Tokyo 152-8551, Japan

*iwaya@wpi-aimr.tohoku.ac.jp

Strontium titanate (SrTiO$_3$) is a perovskite oxide insulator with a band gap of ~3.2 eV, and has been the subject of considerable interest because of its huge potential in applications involving superconductivity[1], photocatalysis[2], ferroelectrics[3], quantum paraelectricity[4] and as a key material for oxide-based electronic devices[5]. Recently, SrTiO$_3$ has attracted greater attention owing to the discovery of metallic and magnetic interfaces between two nonmagnetic perovskite insulators LaAlO$_3$ and SrTiO$_3$[6,7]. In order to explore and design new unique functionalities at these surfaces and interfaces, more attention should be paid to the electronic structure of the substrate and thin film surfaces at a true atomic scale. To date, the quality of thin films during and after deposition has been characterized mainly by monitoring reflection high-energy electron diffraction (RHEED) oscillations and patterns, and an atomic-scale understanding of surface structure of the thin film has been completely lacking. In particular, it would be highly desirable to characterize surface structures associated with the growth process in addition to local defects.

Scanning tunneling microscopy and spectroscopy (STM and STS) are ideal probes for this purpose but no system combining STM with pulsed laser deposition (PLD) capable of stable spectroscopy measurements has been available. We have recently constructed a PLD-STM system which enables us to perform low temperature STM/STS measurements for thin films without exposing their surfaces to air.

In this study, the surface structure of $SrTiO_3$(001) thin films homoepitaxially grown by PLD in step-flow mode was characterized using low temperature STM observations. It was found that one-dimensional (1D) $TiO_x$-based nanostructures were formed on the thin film surface and their density increased with increasing thin film thickness. Most of the 1D nanostructures disappeared after a post-deposition annealing, indicating that this structure is metastable due to the nonequilibrium growth mode. The resulting surface after annealing exhibited similar features to that of a thinner film, having a domain structure with (2×1) and (1×2) reconstructions, but with fewer oxygen-vacancy-type defects. These results imply that the step-flow growth is likely to produce $TiO_x$-rich surface and Ti deficiencies in the film. By the post-deposition annealing, the rich $TiO_x$ would diffuse from the surface into the film to compensate defects associated with Ti vacancies and oxygen vacancies, resulting in the stable surface structure with fewer oxygen vacancies. Thus, STM measurements can provide us with a microscopic picture of surface stoichiometry of thin films originating in the dynamics of the growth process, and can present a new approach for designing functional oxide films.

Figure 1(a) shows typical RHEED intensity oscillations during film growth, exhibiting exponential recovery to the initial intensity level, confirming the step-flow film growth mode. The RHEED pattern shown in Figure 1(b) exhibits sharp streaks with Kikuchi lines, indicating the high quality of the thin film. The streaks between the specular and (01) spots corresponds to a (2×2) reconstruction. Figure 1(c) displays a wide-scale STM image of a

35-unit-cell (u.c.)-thick SrTiO$_3$(001) film showing a clear step and terrace structure with a single unit cell step height (~0.4 nm). The presence of such straight step edges is in distinct contrast to thin films grown in layer-by-layer mode, for which the step edges tend to be rather indented and small islands and pit structures exist on the terraces[8].

Focusing on a smaller scale, we begin to see characteristic 1D nanostructures preferentially grown along the *a* or *b* crystallographic axis. We found that the density of these nanostructures increases with increasing film thickness (Figure 1(d)-(f)). In the 10 u.c. film (Figure 1(d)), only a few such nanostructures exist, and the surrounding surface exhibits a domain structure with a (2×1) and (1×2) reconstruction with many defects imaged as bright spots. The (2×2) streaks observed in the RHEED pattern are possibly due to this reconstruction. We note that a domain structure with (2×1) and (1×2) reconstructions has been observed using high resolution TEM in single crystal SrTiO$_3$(001) prepared at 950-1000°C in much higher oxygen pressure and a complicated TiO$_x$-rich surface structure has been proposed[9]. Therefore, a similar structure might be formed in this 10 u.c. film. The defects observed in the (2×1) and (1×2) reconstructed surface are likely to be mainly oxygen vacancies because we found that the number of defects can be substantially reduced by growing a 10 u.c. film under higher oxygen pressure (1×10$^{-5}$ Torr). A detailed discussion on these defects will be presented elsewhere[10].

Figure 2(a) shows a close-up view of a 1D nanostructure which is composed of two zig-zag chains. These chains align parallel to and out of phase with each other. The width of the nanostructure is ~1.6 nm, corresponding to 4 u.c. A similar structure has been reported on a single crystal surface prepared by sputtering and annealing, and considered to be a TiO$_x$-based nanostructure growing on a TiO$_2$-rich surface based on Auger electron spectroscopy and STM measurements[11].

In order to study the local electronic states in more detail, we performed STS measurements on the 10 u.c. film (Figure 2(b)). The spectra were taken with a resolution of 128 × 128 pixels in a 25 nm × 25 nm region, and we can categorize those spectra into three types depending on the location on the surface - the 1D nanostructure, the surrounding (2×1) and (1×2) reconstructed surface away from defects, and defect sites. The *dI/dV* spectra taken at the 1D nanostructure and the (2×1) and (1×2) surface have a quite similar line shape and no apparent peak structure, but the defect regions (bright spots) show characteristic peaks at around 2.6 - 3.4 V which are distributed over a wide energy range. A representative *dI/dV* spectrum showing the peak structure is shown in Figure 2(b). This result suggests that the 1D nanostructure has a similar stoichiometry to the surrounding (2×1) and (1×2) structures, in which, however, many defects associated with oxygen vacancies exist. Actually, a strong bias dependence was observed in STM images at higher energies in the empty states due to these defects which have a higher density of states than the 1D nanostructure (Figure 2(c)).

As seen in Figure 3(a), the surface of the thickest 100 u.c. film is completely covered with the 1D nanostructures, clearly showing a (6×2) surface reconstruction in the fast Fourier transform (FFT) image (Figure 3(b)). We note that x-ray diffraction (XRD) analysis of this film showed *c*-axis expansion of only 0.001 nm, which means that the film is nearly stoichiometric[12]. These results indicate that, despite nearly stoichiometric growth condition, increasing the film thickness in the step-flow mode results in the growth of more abundant Ti-related complexes on the surface, accompanying with the increase of Ti deficiencies in the underlying film.

To investigate the effects of post-deposition annealing on this surface structure, the 100 u.c. film was further annealed for 1 hour at 1100°C in $P_{O2}$ ~ 1×10$^{-6}$ Torr after the deposition.

Following annealing, the surface structure dramatically changed as shown in Figure 3(c). Most of the 1D nanostructures disappeared and the surrounding surface exhibited (2×1) and (1×2) domains distributed almost at random (Figure 3(d)) with many defects, which is quite similar to the case for the thinner 10 u.c. film (Figure 1(d)). This result suggests that the 1D nanostructure is metastable, probably due to the highly nonequilibrium growth condition.

We also performed STS measurements on this annealed film. The *dI/dV* spectra taken at defect regions show similar peak structures in the empty states, while no peak was found in the (2×1) and (1×2) surfaces away from defects (Figure 4(a)). To perform a more detailed comparison with the 10 u.c. thin film, we plot peak voltage probability distributions for the two films in Figure 4(b). In the 10 u. c. film, the peak voltage is distributed over a wide range of ~0.8 V, with the most common value being ~3.2 V. In contrast, the annealed 100 u.c. film has peaks in a relatively narrow range of higher voltages, and the defect density estimated from STM images is also lower, implying that the distribution of electronic states in the annealed film is more uniform than in the 10 u.c. film.

Taken together, these results suggest a near-surface stoichiometry of thin films depending on thickness and the post-deposition annealing as illustrated in Figure 4(c). In the 10 u.c. film, a few 1D nanostructures and many local defects were observed on the (2×1) and (1×2) reconstructed surface, indicating existence of a few Ti vacancies in the film and many oxygen vacancies in the surface. By increasing film thickness, the density of 1D nanostructure increases, suggesting the increase of $TiO_x$ on the surface and Ti vacancies in the film. After the post-deposition annealing, the $TiO_x$–based 1D nanostructures would diffuse into the underlying layer to be incorporated into the Ti vacancies and oxygen vacancies, and the most stable (2×1) and (1×2) reconstructed surface is formed with fewer oxygen vacancies. Since the subsurface stoichiometry is inaccessible to STM, we performed XRD measurements for the 100 u.c. films before and after the post-deposition annealing, but no clear difference was found between them. Therefore, the drastic annealing effect observed in STM images might occur only near the surface. However, we cannot rule out the possibility that the difference of stoichiometry between these two films is too small to be detected by XRD. Other probe capable of imaging depth profile, e.g. high-resolution TEM, might be able to find the answer.

To date, most attempts to understand various interesting phenomena occurring at the oxide interface have been based on the concept of ideal surface termination. However, our results demonstrate that the surface structure of $SrTiO_3$ thin film grown in the step-flow mode is more complicated than that of the original crystal structure. In addition, numerous defects associated with Ti vacancies and oxygen vacancies exist near the surface, and post-deposition annealing is an effective means of obtaining a stable surface structure with fewer defects. These results reveal that STM can be a powerful probe to characterize the electronic structure of a wide variety of thin films at a true atomic scale and can provide us with new insights into designing novel functional oxides.

In conclusion, we have performed the first investigations of the local surface structure of oxide thin films using low temperature STM. The surface of $SrTiO_3(001)$ homoepitaxial thin films grown in step-flow mode with PLD were found to exhibit $TiO_x$-based 1D nanostructures on a $TiO_x$-rich (2×1) and (1×2) reconstructions, and the density of these nanostructures increased with increasing film thickness, indicating that the step-flow mode tends to produce a $TiO_x$-rich surface. By post-deposition annealing, the metastable 1D nanostructures would diffuse into the film to incorporate the Ti vacancies and oxygen vacancies, and the (2×1) and (1×2) domain structure was formed in the surface with fewer defects. Our data represents that the post-deposition annealing is effective to obtain electronically uniform thin film surfaces.

[Experimental]

The $SrTiO_3$(001) thin films used in this study were homoepitaxially grown on Nb-doped (0.05 wt%) $SrTiO_3$(001) substrates by PLD. The substrate surfaces were prepared by buffered HF etching[13] and subsequent annealing at 1000˚C for 0.5 hour in a partial oxygen pressure $P_{O2}$ of ~ $1 \times 10^{-6}$ Torr. The substrate surface showed a clear step and terrace structure, but no atomically-ordered structure was observed[8]. Thin film growth was performed in step-flow mode at 1100˚C in $P_{O2}$ of ~$1 \times 10^{-6}$ Torr, and was monitored using RHEED. A KrF excimer laser ($\lambda$ = 248 nm) with a repetition rate of 1 Hz was used, and the laser fluence at the target surface was ~2.3 J/cm$^2$. We prepared three samples with thickness of 4 nm (10 u.c.), 14 nm (35 u.c.), and 40 nm (100 u.c.). After film growth, the samples were cooled to room temperature at a rate of ~ 3 K/sec, and transferred to the STM system without exposing the surface to air. All STM measurements were performed at 78 K in ultrahigh vacuum conditions.

[Acknowledgement]

This study was supported by the World Premier Research Institute Initiative promoted by the Ministry of Education, Culture, Sports, Science, and Technology, Japan.

**Figure 1**

(a) RHEED intensity oscillations during SrTiO$_3$(001) homoepitaxial growth at 1100°C in an oxygen partial pressure of 1×10$^{-6}$ Torr.

(b) RHEED pattern of 10 u.c. SrTiO$_3$(001) thin film after growth.

(c) Wide-area constant-current STM image of homoepitaxially-grown SrTiO$_3$(001) thin film (thickness: 35 u.c.) showing a clear step and terrace structure with a single unit cell height (200 nm × 200 nm. Sample-bias-voltage $V_s$ = +2 V, and the set-point tunneling current $I_t$ = 40 pA).

(d) –(f) Thickness dependence of surface structure of SrTiO$_3$(001) thin films (40 nm × 40 nm, $V_s$ = +2 V, $I_t$ = 40 pA). (d) 10 u.c. (e) 35 u.c. (f) 100 u.c.

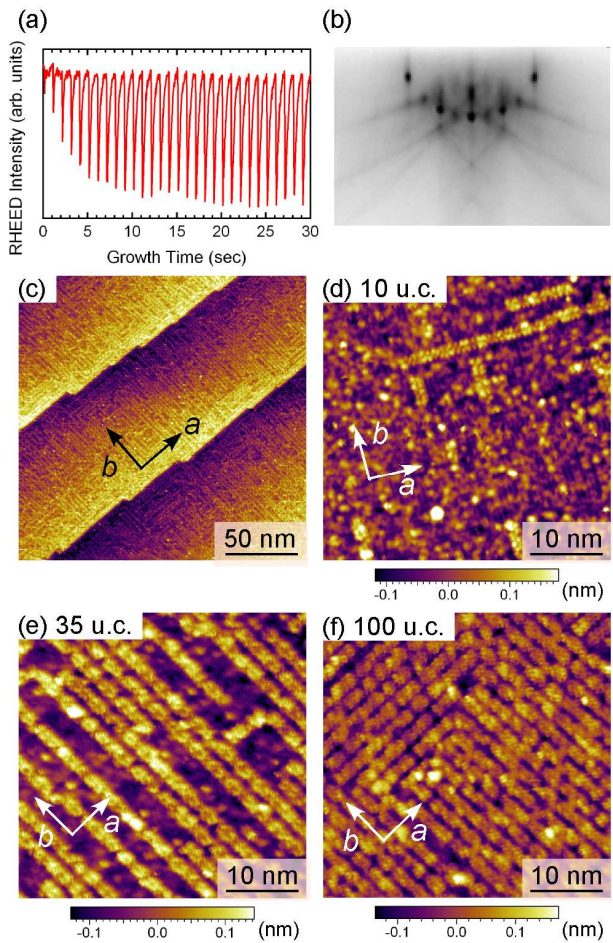

**Figure 2**

(a) Close-up view of the 1D nanostructure. The width of the structure is ~1.6 nm (4 u.c.).

(b) $dI/dV$ spectra measured at three characteristic locations: the 1D nanostructure (○), the surrounding (2×1) and (1×2) surface away from defects (×), and at defect sites (+) (Set point: $V_s = +3.5$ V, $I_t = 40$ pA). Inset: STM image of 10 u.c. film. 25 nm × 25 nm.

(c) Bias-voltage dependent STM images around the 1D nanostructure. $V_s = +2, +3, +3.5$ V. Scan size is 9 nm × 43 nm.

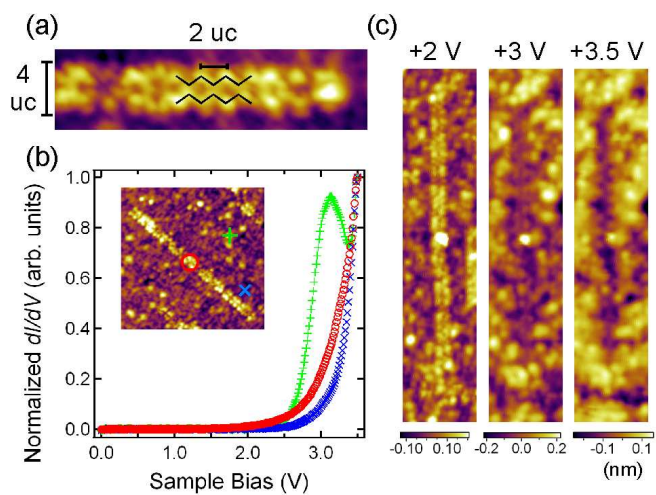

**Figure 3**

(a) Constant-current STM image of 100 u.c. SrTiO$_3$(001) film at $V_s$ = +1 V. $I_t$ = 40 pA. 13 nm × 13 nm.

(b) FFT image corresponding to (a) showing a (6×2) surface reconstruction. The symmetry is slightly distorted due to thermal drift during scanning.

(c) STM image of annealed 100 u.c. SrTiO$_3$(001) film. $V_s$ = +1.5 V. $I_t$ = 40 pA. 25 nm × 25 nm.

(d) Line profile along A-B and C-D in (c) showing a domain structure of (2×1) and (1×2) reconstructions.

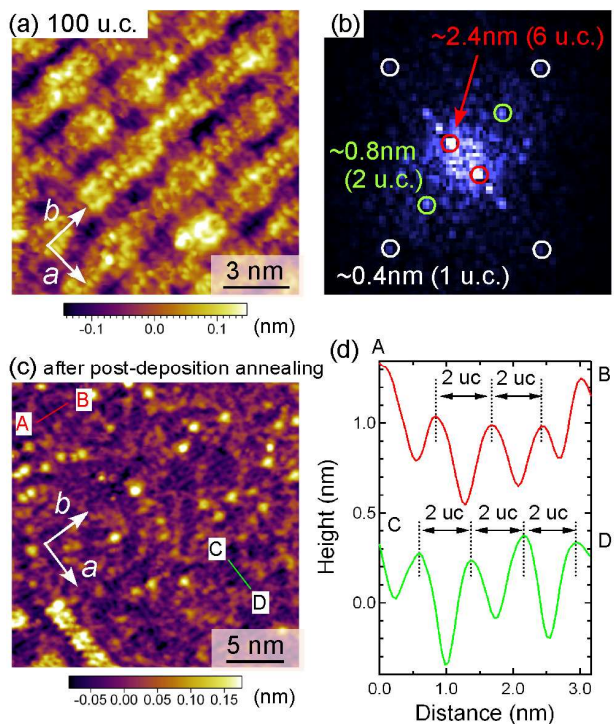

**Figure 4**

(a) Representative $dI/dV$ spectra of 100 u.c. film after post-deposition annealing taken at a defect (+) and far from defects (×) as indicated in the inset. (Set point: $V_s = +3.5$ V, $I_t = 40$ pA). Peak structure associated with the defect is indicated by an arrow. Inset: $dI/dV$ map at $V_s = +2.8$ V (128 × 128 pixels, 20 nm × 20 nm).

(b) $dI/dV$ peak voltage probability distributions at defect sites. Upper: 10 u.c. film. Bottom: 100 u.c. film after post-deposition annealing. The defect density estimated from STM images is also shown.

(c) Schematic of near-surface structure of $SrTiO_3$(001) thin films depending on thickness and the post-deposition annealing. Hatched regions are inaccessible to STM.

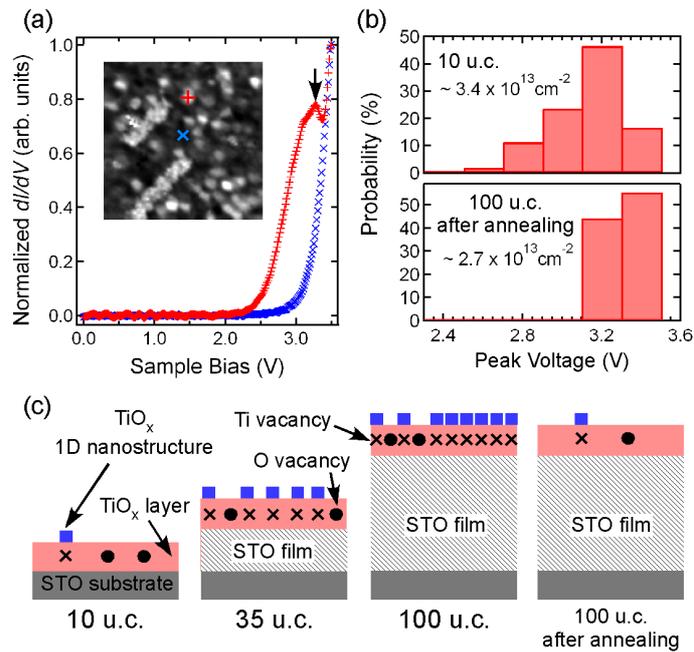